\documentclass[preprintnumbers,nofootinbib,superscriptaddress,10pt]{revtex4}
\usepackage[hypertex]{hyperref}
\usepackage[dvipdfmx]{graphicx} 
\usepackage{amsmath,amssymb,amsfonts,bm} 

\def\a{\alpha} \def\b{\beta} \def\g{\gamma}  \def\d{\delta} \def\D{\Delta}    \def\th{\theta}   \def\l{\lambda}                  

  \def\nn{\nonumber}

\usepackage{color}

\begin{document}

\title{\large 
Unified description of sum rules and duality between CP phases and unitarity triangles through third-order rephasing invariants}

\preprint{STUPP-25-286}

\author{Masaki J. S. Yang}
\email{mjsyang@mail.saitama-u.ac.jp}
\affiliation{Department of Physics, Saitama University, 
Shimo-okubo, Sakura-ku, Saitama, 338-8570, Japan}
\affiliation{Department of Physics, Graduate School of Engineering Science,
Yokohama National University, Yokohama, 240-8501, Japan}



\begin{abstract} 

In this paper, we demonstrate that products of third-order rephasing invariants $V_{\alpha i} V_{\beta j} V_{\gamma k} / \det V$ of flavor mixing matrix $V$ reproduce  all the nine angles of unitarity triangles and all the CP phases in the nine parameterizations of $V$.
The sum rules relating the CP phases and angles are also decomposed into terms of these rephasing invariants.
Furthermore, through ninth-order invariants, 
these fourth- and fifth-order invariants become equivalent, 
which can be regarded as a certain duality. 
For the phase matrix $\Delta$ and the angle matrix $\Phi$,
$\Delta \pm \Phi$ are expressed in terms of even-permutations X and odd-permutations $\Psi$ of third-order invariants.
As a result, these are represented by the two concise matrix equations 
$\Phi = \Psi - {\rm X}$ and $\Delta =  \Pi' - \Psi - {\rm X}$.

\end{abstract} 

\maketitle

\section{Introduction}

The flavor mixing matrix provides an essential formalism of flavor physics. 
Since its CP phase is intimately related to the origin of the matter-antimatter asymmetry, 
it is one of the key observables for particle physics in the near future.
In order to establish a theoretical basis, rephasing invariants of the mixing matrix \cite{Wu:1985ea, Bernabeu:1986fc, Gronau:1986xb, Branco:1987mj,  Bjorken:1987tr, Nieves:1987pp,  Botella:1994cs, Branco:1999fs, Jenkins:2007ip} have been studied
to define CP-violating quantities  and to elucidate their physical significance.
The fourth-order rephasing invariants, represented by the Jarlskog invariant \cite{Jarlskog:1985ht}, have been widely employed in the analysis of unitarity triangles
\cite{Wu:1994di, Hocker:2006xb, Xing:2009eg, Harrison:2009bz, Antusch:2009hq, Frampton:2010ii, Dueck:2010fa, Frampton:2010uq,Li:2010ae,Qin:2010hn, Zhou:2011xm,Qin:2011ub,Zhang:2012bk, Zhang:2012ys, Li:2012zxa, He:2013rba, He:2016dco, Xing:2019tsn, Kaur:2023ypg,  Harrison:2025rkp}. 
However, they are insufficient for providing a unified understanding of the CP phases in the mixing matrix. 
Indeed, CP phases introduced in the nine Euler-angle-like parameterizations \cite{Fritzsch:1997st} appear to take unrelated and scattered values at first glance.

Recently, some studies have begun focusing on rephasing invariants that explicitly incorporate the determinant of the mixing matrix $V$ \cite{Yang:2025hex}. 
Unlike the Jarlskog-type quartic invariants, these new invariants involving $\det V$ provide a concise  framework that can express the CP-violating phases.
A specific relation $\d_{\rm PDG} + \d_{\rm KM} = \pi - \a + \g$ 
presented a nontrivial connection between the CP phases $\delta_{\rm PDG}, \delta_{\rm KM}$ and the angles $\alpha, \gamma$ of the unitarity triangle \cite{Yang:2025ftl}.
Furthermore, the set of nine formulae for the CP phases $\delta^{(\alpha i)} = \arg [ { V_{\alpha 1} V_{\alpha 2} V_{\alpha 3} V_{1i} V_{2i} V_{3i} / V_{\alpha i }^{3} \det V } ]$ and its generalized sum rules 
 \cite{Yang:2025vrs} reveal an underlying regularity between chaotic CP phases and unitarity triangles. 
Historically, in particle physics, such regularities have often indicated the existence of more fundamental physical quantities.

In this study, we present six third-order rephasing invariants with determinant \cite{Kuo:2005pf, Yang:2025law,  Luo:2025wio}
reproduce  all nine CP angles of the unitarity triangles and all CP phases in the nine different parameterizations.
The sum rules relating the CP angles and  phases are also decomposed into expressions of these fewer invariants. 
Furthermore, through ninth-order invariants, we demonstrate the existence of a duality between the nine CP phases and the nine angles.

This paper is organized as follows. 
The next section gives a rephasing invariant formalism of unitarity triangles and CP phases of the  mixing matrix. 
In Sec.~3, we demonstrate decomposition of sum rules by third-order rephasing invariants and a duality between the CP phases and the unitarity triangles. 
The final section is devoted to a summary. 

\section{Nine angles of unitarity triangles and nine CP phases in parameterizations of mixing matrix}

Here, we first review the general sum rules between nine angles of the unitarity triangles and CP phases in the nine Euler-angle-like parameterizations of the mixing matrix.  
The observed values of the CKM matrix from the latest UTfit are \cite{UTfit:2022hsi}, 
\begin{align}
\sin \th_{12} &= 0.22519 \pm 0.00083 \, ,  ~~~ \sin \th_{23} = 0.04200 \pm 0.00047 \, ,  \nn \\
\sin \th_{13} &= 0.003714 \pm 0.000092 \, ,  ~~~ \d = 1.137 \pm 0.022  = 65.15^{\circ} \pm1.3^{\circ}  \, ,
\end{align}
and the angles of the unitarity triangles for the best-fit values are
\begin{align}
\a = \arg \left [ - { V_{td}^{} V_{tb}^{*} \over V_{ud}^{} V_{ub}^{*} } \right ] = 92.40^{\circ} , ~~ 
\b = \arg \left [ - { V_{cd}^{} V_{cb}^{*} \over V_{td}^{} V_{tb}^{*} }  \right ] = 22.49^{\circ} , ~~ 
\g = \arg \left [ - { V_{ud}^{} V_{ub}^{*} \over V_{cd}^{} V_{cb}^{*} }  \right ] = 65.11 ^{\circ} .
\end{align}
Although the six unitarity triangles have a total of 18 angles, each pair is equivalent, so there are only nine distinct angles. They have been organized and defined as \cite{Harrison:2009bz}
\begin{align}
\Phi 
\equiv 
\begin{pmatrix}
\arg (-\Pi_{ud}^*) & \arg (-\Pi_{us}^*) & \arg (-\Pi_{ub}^*) \\
\arg (-\Pi_{cd}^*) & \arg (-\Pi_{cs}^*) & \arg (-\Pi_{cb}^*) \\
\arg (-\Pi_{td}^*) & \arg (-\Pi_{ts}^*) & \arg (-\Pi_{tb}^*) \\
\end{pmatrix}
= 
\begin{pmatrix}
 1.054^{\circ} & 22.49^{\circ} & 156.46^{\circ} \\
 64.09^{\circ} & 92.40^{\circ} & 23.50^{\circ} \\
 114.85^{\circ} & 65.11^{\circ} & 0.0370^{\circ} \\
 \end{pmatrix} , 
\end{align}
where $\Pi_{\alpha i} \equiv V_{\beta j} V_{\beta k}^* V_{\gamma k} V_{\gamma j}^* $
and 
\begin{align}
\Pi = 
\begin{pmatrix}
  V_{tb} V_{ts}^* V_{cs} V_{cb}^*
& V_{td} V_{tb}^* V_{cb} V_{cd}^*
& V_{ts} V_{td}^* V_{cd} V_{cs}^*  \\
   V_{ub} V_{us}^* V_{ts} V_{tb}^*
& V_{ud} V_{ub}^* V_{tb} V_{td}^*
& V_{us} V_{ud}^* V_{td} V_{ts}^*   \\
   V_{cb} V_{cs}^* V_{us} V_{ub}^*
& V_{cd} V_{cb}^* V_{ub} V_{ud}^*
& V_{cs} V_{cd}^* V_{ud} V_{us}^* 
\end{pmatrix} . 
\end{align}
In particular, the elements of the second column are the angles $\beta, \alpha, \gamma$, respectively. 
Hereafter, for better readability, the quark flavor labels are assigned as $u,c,t = 1,2,3$ and $d,s,b = 1,2,3$.
This matrix is constructed so that the sum over each row and each column is equal to $180^\circ$, 
\begin{align}
\sum_{i} \Phi_{ij} = \sum_{j} \Phi_{ij} = \pi \, .  \label{phipi}
\end{align}

On the other hand,  Fritzsch and Xing have pointed out that there exist nine inequivalent Euler-angle parameterizations of the mixing matrix \cite{Fritzsch:1997st}. 
Since these matrix elements have trivial phases in certain row $\alpha$ and column $i$ with $V_{\alpha j}, V_{\beta i} \in \mathbb{R}$, we denote their phases by $\delta^{(\alpha i)}$ to distinguish between them.
The structure of  CP phases in such a parameterization $V^{(\alpha i)}$ is determined by the following conditions 
\begin{align}
\arg  \det V^{(\a i )} = - \d^{(\a i )} \, , ~
\arg V_{\a 1}^{(\a i)} = \arg V_{\a 2}^{(\a i)} = \arg V_{\a 3}^{(\a i)} = \arg V_{1i}^{(\a i)} = \arg V_{2 i }^{(\a i )} = \arg V_{3 i}^{(\a i )} =  0 ~ {\rm or} ~ \pi \, . 
\end{align}
Since there is a duplication for the element $V_{\alpha i}^{(\alpha i)}$, there are effectively six independent conditions, and the phase $\pi$ appears either zero or two times. 

To determine these phases $\delta^{(\alpha i)}$, 
we consider rephasing transformations from  $V^{(\alpha i)} $ to the mixing matrix $V$ in an arbitrary basis 
\begin{align}
V = \Psi_{L} V^{(\a i )}  \Psi_{R} ^{\dagger} \, , ~~
V_{\b j} = e^{i \g_{L \b}} V^{(\a i)}_{\b j} e^{- i \g_{R j}} \, , 
\end{align}
where $(\Psi_{L})_{\alpha\beta} = e^{i \gamma_{L \alpha}} \delta_{\alpha \beta}$ and $(\Psi_{R})_{ij} = e^{i \gamma_{R i}} \delta_{ij}$ are diagonal phase matrices.
Comparing the determinants of the two representations,
one finds $\arg \det V = - \delta^{(\alpha i)} + \sum_{\beta, j} (\gamma_{L \beta} - \gamma_{R j})$ 
and the CP phases $\delta^{(\alpha i)}$ \cite{Yang:2025vrs}
\begin{align}
\d^{(\a i )} & =  (\g_{L1} + \g_{L2} + \g_{L3} - \g_{R1} - \g_{R2} - \g_{R3}) - \arg  \det V \nn \\
& = 
\arg [V_{\a 1} V_{\a 2} V_{\a 3} V_{1i} V_{2i} V_{3i} / V_{\a i}^{3}] - \arg  \det V
= \arg \left[ { V_{\a 1} V_{\a 2} V_{\a 3} V_{1i} V_{2i} V_{3i}  \over V_{\a i }^{3} \det V } \right] . 
\end{align}
More specifically, the two factors $V_{\alpha i}$  in the numerator cancel those in the denominator, 
\begin{align}
\d^{(11)} & = \arg \left[ { V_{12} V_{13}  V_{21} V_{31}  \over V_{11} \det V } \right]  , ~~ 
\d^{(12)} = \arg \left[ { V_{11} V_{13}  V_{22} V_{32}  \over V_{12} \det V} \right] , ~~
\d^{(13)} = \arg \left[ { V_{11} V_{12}  V_{23} V_{33}  \over V_{13} \det V} \right]  , \nn \\
\d^{(21)} & = \arg \left[ { V_{22} V_{23} V_{11}  V_{31}  \over V_{21} \det V} \right]  , ~~ 
\d^{(22)} = \arg \left[ {V_{21} V_{23}V_{12}  V_{32}   \over V_{22} \det V} \right]  , ~~
\d^{(23)} = \arg \left[ { V_{21} V_{22} V_{13} V_{33}  \over V_{23} \det V} \right]  , \nn \\
\d^{(31)} & = \arg \left[ { V_{32} V_{33} V_{11} V_{21}   \over V_{31}\det V} \right]  , ~~ 
\d^{(32)} = \arg \left[ { V_{31} V_{33}  V_{12} V_{22}  \over V_{32} \det V} \right]  , ~~
\d^{(33)} = \arg \left[ { V_{31} V_{32} V_{13} V_{23}  \over V_{33} \det V} \right] . 
\end{align}
These constitute a necessary and sufficient set of all ``irreducible fifth-order'' invariants which cannot be decomposed into second- and third-orders. 
The well-known examples are the phase of PDG parameterization $\delta_{\rm PDG}$ and of the original Kobayashi--Maskawa parameterization $\delta_{\rm KM}$, 
\begin{align}
\d^{(11)} = \pi - \d_{\rm KM} \, , ~~ \d^{(13)} = \d_{\rm PDG} \, . 
\label{dKMPDG}
\end{align}

From such expressions, some nontrivial relations are expected between $\Phi_{\alpha i}$ and $\delta^{(\alpha i)}$. 
In fact, the following exact sum rule between the phases and the angles has been discovered \cite{Yang:2025ftl}
\begin{align}
\d_{\rm PDG} + \d_{\rm KM} = \pi - \a + \g  \, , ~~~ 
\d^{(1 1)} - \d^{(1 3)}  = \Phi_{22} - \Phi_{32} \, . 
\label{KMsum}
\end{align}
A more general sum rules are found to be \cite{Yang:2025vrs}
\begin{align}
 \D T^{2} - \D T = T^{2}  \Phi - T \Phi  \, , ~~~ T^{2} \D - T \D = \Phi T^{2} - \Phi T \, , 
 \label{oldsumrules}
\end{align}
where the phase matrix $\Delta$ and the transfer matrix $T$ are defined as 
\begin{align}
\D = 
\begin{pmatrix}
\d^{(11)} & \d^{(12)} & \d^{(13)} \\
\d^{(21)} & \d^{(22)} & \d^{(23)} \\
\d^{(31)} & \d^{(32)} & \d^{(33)} 
\end{pmatrix} , ~
T = 
\begin{pmatrix}
 0 & 0 & 1 \\
 1 & 0 & 0 \\
 0 & 1 & 0 \\
\end{pmatrix} . 
\end{align}
Many of these nontrivial relations suggest the existence of more fundamental underlying quantities.  
In the next section, we show that all nine CP angles and all nine CP phases are derived by third-order  rephasing invariants.

\section{Third-order rephasing invariants with determinant}

In order to express these CP-violating observables by fewer invariants, we employ the six rephasing invariants  as follows \cite{Kuo:2005pf, Yang:2025law, Luo:2025wio}. 
\begin{align}
\chi_{1} &= \arg \left[ \frac{ V_{11} V_{22} V_{33} }{  \det V } \right] 
= - 0.0019^{\circ} \, , ~ 
\chi_{2} = \arg \left[ \frac{  V_{12} V_{23} V_{31} }{  \det V } \right] = - 22.45^{\circ} , ~ 
\chi_{3} = \arg \left[ \frac{  V_{13} V_{21} V_{32} }{ \det V } \right]
= -64.06^{\circ} \, ,  \nn \\
\psi_{1}& = \arg \left[ - \frac{V_{11} V_{23} V_{32} }{  \det V } \right] 
= 1.05^{\circ} \, ,  ~
\psi_{2} = \arg \left[ - \frac{  V_{12} V_{21} V_{33} }{ \det V } \right]
 = 0.035^{\circ} \, , ~
\psi_{3} =  \arg \left[ - \frac{  V_{13} V_{22} V_{31} }{ \det V } \right]
= 92.40^{\circ}  \, .  \nn
\end{align}
The arguments of these third-order invariants correspond to the nontrivial phases of the matrix elements $V^{0}_{\alpha i}$ in the PDG parametrization; 
\begin{align}
\chi_{1} = \arg V_{22}^{0} \, , ~~ \chi_{2} = \arg V_{31}^{0} \, , ~&~  \chi_{3} = \arg V_{13}^{0} + \arg V_{21}^{0} + \arg V_{32}^{0} \, , ~~ \nn \\
\psi_{1} = \arg V_{32}^{0} + \pi \, , ~~ \psi_{2} =  \arg V_{21}^{0} + \pi \, , ~&~  \psi_{3} = \arg V_{13}^{0} + \arg V_{22}^{0} + \arg V_{31}^{0} + \pi  \, . 
\end{align}
By assigning the $\pm$ sign to even and odd permutations, 
all angles become almost acute, and the signs are reflected as well. 
The sum of each of the three angles is equal to 
\begin{align}
\psi_{1} + \psi_{2} + \psi_{3} =  \arg \left[ - \frac{ \prod_{\a, i} V_{\a i} }{  (\det V)^{3} } \right] 
=  \chi_{1} + \chi_{2} + \chi_{3} + \pi = 93.49^{\circ} \, .  
\label{detV3}
\end{align}
This is a ninth-order invariant constructed from all the matrix elements $V_{\alpha i}$, without involving any complex conjugates. 
If these two sums are regarded as constraints, the degrees of freedom among  $\psi_i$ and $\chi_i$ can be four.  

To express $\Phi$ using $\psi_{i}$ and $\chi_{i}$, we define the following matrix:
\begin{align}
{\rm X } =
\begin{pmatrix}
\chi_{1} & \chi_{2} & \chi_{3}    \\
\chi_{3} & \chi_{1} & \chi_{2}  \\
\chi_{2} & \chi_{3} & \chi_{1} \\
\end{pmatrix},  ~~
\Psi = 
\begin{pmatrix}
\psi_{1} & \psi_{2} & \psi_{3} \\
\psi_{2} & \psi_{3} & \psi_{1} \\
 \psi_{3} & \psi_{1} & \psi_{2} \\
\end{pmatrix} . 
\end{align}
From this, it immediately follows that
\begin{align}
\Psi - {\rm X} = 
\begin{pmatrix}
\psi_{1} - \chi_{1} & \psi_{2} - \chi_{2} & \psi_{3} - \chi_{3} \\
\psi_{2} - \chi_{3} & \psi_{3} - \chi_{1} & \psi_{1} - \chi_{2}\\
\psi_{3} - \chi_{2} & \psi_{1} - \chi_{3} & \psi_{2} - \chi_{1}  \\
\end{pmatrix}
 = \Phi \, . 
\label{PXP}
\end{align}
The subsequent letter~\cite{Yang:2025dhm} shows that these arguments of the invariants $\Psi$, ${\rm X}$, and $\Phi$ correspond to the angles of an alternative set of unitarity triangles obtained from the inversion formula $U^{\dagger} = U^{-1}$. Therefore, the sum rules between CP phases and angles presented below are expected to be understood geometrically by compositions of the new unitarity triangles. 

Since the second column corresponds to the unitarity triangle angles, 
$\psi_{2} - \chi_{2} = \b \, , \psi_{3} - \chi_{1} = \a \, , \psi_{1} - \chi_{3} = \g$ hold. 
Moreover, since the three angles $\chi_{1} \, , \psi_{1} \, , \psi_{2}$ are smaller than about $1^\circ$, it is also useful to regard these angles as deviations associated with $\a, \b, \g$, 
\begin{align}
 \chi_{1} = -  d \a \, , ~~ \chi_{2} &= d \b  - \b \, , ~~ \chi_{3} = d \g - \g \, , \nonumber \\
\psi_{1} = d \g \, , ~~ \psi_{2} &= d \b \, ,  ~~ \psi_{3} = \a - d \a \, , 
\label{dabg}
\end{align}
where $d$ symbolically denotes small differences. 
Then, the nine CP angles are
\begin{align}
\Phi = 
\begin{pmatrix}
\psi_{1} - \chi_{1} & \psi_{2} - \chi_{2} & \psi_{3} - \chi_{3} \\
\psi_{2} - \chi_{3} & \psi_{3} - \chi_{1} & \psi_{1} - \chi_{2}\\
\psi_{3} - \chi_{2} & \psi_{1} - \chi_{3} & \psi_{2} - \chi_{1}  \\
\end{pmatrix}
 = 
\begin{pmatrix}
d \a +  d\g & \b & \a+ \g - d\a  - d \g \\
\g + d \b - d \g & \a &\b - d \b + d \g \\ 
\a + \b - d \a - d \b & \g & d \a + d \b\\
\end{pmatrix} , 
\end{align}
and the sum of each row and column clearly reduces to Eq.~(\ref{phipi}), $\alpha + \beta + \gamma = \pi$. 
The three small quantities in the standard PDG parameterization are
\begin{align}
\chi_{1} \simeq - s_{12} s_{13} s_{23} \sin \delta \sim \l^{6} \, ,   ~~
\psi_{2} \simeq \frac{ s_{13} s_{23} }{s_{12}} \sin \delta \sim \l^{4} \, , ~~
\psi_{1} \simeq \frac{ s_{12} s_{13} }{s_{23}} \sin \delta \sim \l^{2} \, . 
\end{align}
Although these do not contain any new information about the CP phases,  each is a small quantity of even order in $\lambda$, and some of them can be neglected for calculations of $O (\lambda^{2n})$. 

It is evident that the phase matrix $\Delta_{\a i} = \delta^{(\a i )}$ can be expressed in terms of the six invariants and the nine CP angles.  
To obtain a representation of $\D$ in which $\pi$ does not explicitly appear, we can only use $\Phi$ and $\Psi$.  
Furthermore, since $\det V$ appears in the denominator of $\D_{\a i}$, the only possibility is to add $\Phi_{\a i }$ and $\psi_i$.  
Among the three possibilities of $\psi_{i}$, there are two cases in which the denominator of $\D_{\a i }$ does not overlap, allowing them to be written as two matrix equations.

As a result, the phase matrix $\Delta$ are expressed using only $\Phi$ and $\Psi$ as follows; 
\begin{align}
\D =  \Psi  T + T^{T} \Phi T  = \Psi T^T + T \Phi T^{T} \, .   
\label{DPsiPhi}
\end{align}
Alternatively, by substituting Eq.~(\ref{PXP}) into Eq.~(\ref{DPsiPhi}),
\begin{align}
\D &= \Psi T + T^{T } ( \Psi - {\rm X}) T  = \Psi T^T + T ( \Psi - {\rm X}) T^{T} \, .  
\label{DPX}
\end{align}
The two equations in Eq.~(\ref{DPX}) are equivalent with respect to ${\rm X}$ and $\Psi$, 
from the transformation properties under $T$
\begin{align}
 T {\rm X} T^{T} = T^{T} {\rm X} T  = {\rm X} \, ,  ~~~ 
 T \Psi T = T^{T} \Psi T^{T}  = \Psi \, , 
\end{align}
and  $T^{2} = T^{T} = T^{-1}$. 
Eqs.~(\ref{PXP}) and (\ref{DPX}) are regarded as the fundamental equations. 

Explicitly, $\D$ is represented as
\begin{align}
\D & = 
\begin{pmatrix}
 -\chi _1+\psi _2+\psi _3 & -\chi _2+\psi _3+\psi _1 & -\chi _3+\psi _1+\psi _2 \\
 -\chi _3+\psi _3+\psi _1 & -\chi _1+\psi _1+\psi _2 & -\chi _2+\psi _2+\psi _3 \\
 -\chi _2+\psi _1+\psi _2 & -\chi _3+\psi _2+\psi _3 & -\chi _1+\psi _3+\psi _1 \\
\end{pmatrix} \label{Deltaij} \nn
\\ & = 
\begin{pmatrix}
 \alpha +d \beta  & \alpha + \beta - d \alpha  - d \beta + d \gamma & \gamma + d \beta  \\
 \alpha +\gamma - d \alpha & d \alpha  + d \beta + d \gamma  & \alpha +\beta - d \alpha  \\
 \beta + d \gamma & \alpha + \gamma - d \alpha  + d \beta - d \gamma  & \alpha + d \gamma  \\
\end{pmatrix} .
\end{align}
Thus, although these fifth-order invariants are apparently irreducible, they are expressed as a product of three third-order invariants.

The sum of each row and column is
\begin{align}
\sum_{i} \d^{(ij)} = \sum_{j} \d^{(ij)} =  \pi + \a  - d \a + d \b + d \g
= \pi + \psi_{1} + \psi_{2} + \psi_{3} = 273.49^{\circ} \, ,  \label{24}
\end{align}
which reduces to Eq.~(\ref{detV3}). 
For $\Delta$ and $\Phi$, the sum conditions imply five independent constraints over the rows and columns because the sixth becomes trivial.  
If these five constraints are imposed among the nine phases and angles, the essential degrees of freedom are four, which is consistent.

\subsection*{Decomposition of sum rules}

From these definitions, sum rules between the CP phases and angles 
are described by more fundamental invariants.  
First, let us examine how the specific sum rule (\ref{KMsum}) is expressed in terms of invariants $\psi_{i}$ and $\chi_{i}$.  From Eqs.~(\ref{PXP}) and (\ref{Deltaij}),
\begin{align}
\Phi_{22} = \a = \psi_{3} - \chi_{1} \, , ~ & ~   \Phi_{32} = \g = \psi_{1} - \chi_{3} \, , 
\\
\d^{(11)} = \pi - \d_{\rm KM} = 
 -\chi _1+\psi _2+\psi _3 \, , ~&~ \d^{(13)} = \d_{\rm PDG}  =  -\chi _3+\psi _1+\psi _2 \, . 
\end{align}
Therefore, the sum rule is rewritten in the following form 
\begin{align}
\Phi_{22} - \Phi_{32} = \a - \g = \psi _3 -\chi _1 - \psi _1 + \chi _3  =
\pi - \d_{\rm KM} - \d_{\rm PDG} =  \d^{(11)} - \d^{(13)} \, . 
\end{align}
An expression involving only a third-order invariant is 
\begin{align}
\d_{\rm PDG} - \g =  \psi _2 = \pi - \d_{\rm KM} -  \a \, . 
\label{28}
\end{align}

To consider such a description of the sum rules, we confirm the previous results  using $T$.
By multiplying Eq.~(\ref{DPsiPhi}) from the right by $T$ and $T^{T}$ gives
\begin{align}
\D T  = \Psi + T \Phi \, ,  ~ \D T^{T} =  \Psi + T^{T} \Phi \, .  
\label{DTPsi1}
\end{align}
Eliminating $\Psi$, one immediately obtains $\D T - T \Phi = \D T^{T} - T^{T} \Phi$ which  
coincides to Eq.~(\ref{oldsumrules}). 
Conversely, solving for $\Psi$ yields 
\begin{align}
\Psi = \D T -  T \Phi  = \D T^{T} - T^{T} \Phi \, .  
\label{30}
\end{align}
Although the two expressions are superficially inequivalent, 
\begin{align}
\begin{pmatrix}
\psi_{1} & \psi_{2} & \psi_{3} \\
\psi_{2} & \psi_{3} & \psi_{1} \\
 \psi_{3} & \psi_{1} & \psi_{2} \\
\end{pmatrix}
 & =
\begin{pmatrix}
\d^{(12)} & \d^{(13)}  & \d^{(11)} \\
\d^{(22)} & \d^{(23)} & \d^{(21)} \\
\d^{(32)} & \d^{(33)} & \d^{(31)} 
\end{pmatrix} 
- 
\begin{pmatrix}
\Phi_{31} & \Phi_{32} & \Phi_{33} \\
\Phi_{11} & \Phi_{12} & \Phi_{13} \\
\Phi_{21} & \Phi_{22} & \Phi_{23} \\
\end{pmatrix}  \nn
\\ & = 
\begin{pmatrix}
\d^{(13)} & \d^{(11)} & \d^{(12)} \\
\d^{(23)} & \d^{(21)} & \d^{(22)} \\
\d^{(33)} & \d^{(31)} & \d^{(22)}
\end{pmatrix} 
-
\begin{pmatrix}
\Phi_{21} & \Phi_{22} & \Phi_{23} \\
\Phi_{31} & \Phi_{32} & \Phi_{33} \\
\Phi_{11} & \Phi_{12} & \Phi_{13} \\
\end{pmatrix} ,  \label{31}
\end{align}
the essence of the sum rules is that they constitute an algebraically trivial identity for $\chi_{i}$ and $\psi_{i}$. 
Eq.~(\ref{28}) corresponds to the 1-2 element of Eq.~(\ref{31}).

There exist other equivalent representations for $\Psi$. 
Multiplying Eq.~(\ref{DPsiPhi}) from the left by $T$ and $T^{T}$, 
\begin{align}
 T \D = \Psi + \Phi T \, , ~~ T^{T} \D = \Psi + \Phi T^{T}  \, . 
 \label{TDPsi2}
\end{align}
Eliminating $\Psi$, 
we obtain the other sum rules in Eq.~(\ref{oldsumrules}),  $ T^{2} \D - T \D = \Phi T^{2} -  \Phi T $.
Similarly, rewriting these equations in terms of $\Psi$ yields
\begin{align}
 \Psi =  T \D - \Phi T =  T^{T} \D - \Phi T^{T} \, . 
 \label{37}
\end{align}
However, it corresponds to the same conditions as Eq.~(\ref{31}), 
and does not introduce any additional constraints. 
This is because Eqs.~(\ref{30}) and (\ref{37}) are equivalent under $\Psi = T \Psi T$.

Such relations can also be expressed by ${\rm X}$ instead of $\Psi$.  
Let us note the following fact 
\begin{align}
1 + T + T^{2} = 
\begin{pmatrix}
1 & 1 & 1 \\
1 & 1 & 1 \\
1 & 1 & 1 \\
\end{pmatrix} \, . 
\end{align}
From this fact, Eq.~(\ref{phipi}) is rewritten as
\begin{align}
 (1 + T + T^{2}) \Phi  =  (1 + T + T^{2}) \pi \, . 
 \label{TPi}
\end{align}
By substituting this together with $\Phi = \Psi - {\rm X}$ into Eq.~(\ref{DTPsi1}), 
it is evident that
\begin{align}
\D T = \Psi + T \Phi = {\rm X} +   (1 + T + T^{2}) \pi  - T^{T}  \Phi  \, , ~ 
\D T^{T} = \Psi + T^{T} \Phi =  {\rm X} +  (1 + T + T^{2}) \pi - T \Phi  \, .
\end{align}
Eliminating ${\rm X}$ reduces the expression (\ref{oldsumrules}). On the other hand, solving for ${\rm X}$ yields 
\begin{align}
{\rm X} +  (1 + T + T^{2}) \pi = \D T + T^{T} \Phi = \D T^{T} + T \Phi \, . 
\end{align}
In the same way, it is expressed as 
\begin{align}
\begin{pmatrix}
\chi_{1} & \chi_{2} & \chi_{3}    \\
\chi_{3} & \chi_{1} & \chi_{2}  \\
\chi_{2} & \chi_{3} & \chi_{1} \\
\end{pmatrix}
+
\begin{pmatrix}
\pi & \pi & \pi \\
\pi & \pi & \pi \\
\pi & \pi & \pi \\
\end{pmatrix} 
 & =
\begin{pmatrix}
\d^{(12)} & \d^{(13)}  & \d^{(11)} \\
\d^{(22)} & \d^{(23)} & \d^{(21)} \\
\d^{(32)} & \d^{(33)} & \d^{(31)} 
\end{pmatrix} 
+ 
\begin{pmatrix}
\Phi_{21} & \Phi_{22} & \Phi_{23} \\
\Phi_{31} & \Phi_{32} & \Phi_{33} \\
\Phi_{11} & \Phi_{12} & \Phi_{13} \\
\end{pmatrix}  \nn
\\ & = 
\begin{pmatrix}
\d^{(13)} & \d^{(11)} & \d^{(12)} \\
\d^{(23)} & \d^{(21)} & \d^{(22)} \\
\d^{(33)} & \d^{(31)} & \d^{(22)}
\end{pmatrix} 
+ 
\begin{pmatrix}
\Phi_{31} & \Phi_{32} & \Phi_{33} \\
\Phi_{11} & \Phi_{12} & \Phi_{13} \\
\Phi_{21} & \Phi_{22} & \Phi_{23} \
\end{pmatrix} . 
\end{align}

In this situation, the conditions might seem to be doubled with $\psi_{i}$ and $\chi_{i}$.
However, for example, the 1-2 element is equivalent to Eq.~(\ref{28}), 
\begin{align}
\chi_{2} + \pi & = \d^{(13)}  + \Phi_{22} = \d^{(11)} +  \Phi_{32}  = \delta_{\rm PDG} + \a = \pi - \d_{\rm KM} +  \g  \, . 
\end{align}
It corresponds to a simple transposition of terms between equalities as 
\begin{align}
 \arg \left[- \frac{  V_{12} V_{23} V_{31} }{  \det V } \right]  
= 
\arg \left[ { V_{11} V_{12}  V_{23} V_{33}  \over V_{13} \det V} \right]  
+ 
\arg \left[ - { V_{13}  V_{31} \over V_{11} V_{33} } \right]
= 
 \arg \left[ { V_{12} V_{13}  V_{21} V_{31}  \over V_{11} \det V } \right]  
+ 
 \arg \left[ -  {V_{23} V_{11} \over V_{21} V_{13}} \right]   . 
\end{align}
The reason is that the fundamental equations consist  of Eqs.~(\ref{PXP}) and (\ref{DPX}), 
and the rewriting from $\Psi$ into X relies on Eqs.~(\ref{PXP}), $\Psi = {\rm X} + \Phi$. 
While the representation is doubled with respect to $\Psi$ and X, the sum rules for $\Phi$ and $\D$ stay the same. 
On the other hand, since there are four ways to construct a third-order invariant from
products of a fifth-order invariant and fourth-order invariants, such representations exist in 36 forms for the nine fifth-order invariants, and the counting is consistent.

Finally, the following also holds, 
\begin{align}
\Phi  (1 + T + T^{2})  =  (1 + T + T^{2}) \pi \, ,
\end{align}
and it immediately follows from Eq.~(\ref{TDPsi2}) that
\begin{align}
T \D =   \Psi + \Phi T =  (1 + T + T^{2}) \pi + {\rm X} - \Phi T^{T} \, , ~~ 
T^{T} \D = \Psi + \Phi T^{T} =  (1 + T + T^{2}) \pi + {\rm X} - \Phi T \, .  
\end{align}
As anticipated, these conditions are also not independent. 
Since these are decompositions of the sum rules into the six fundamental invariants, it will be free of redundancy and easier to handle.

\subsection*{Duality between CP phases and unitarity triangles}

As we have seen, the nine angles are equivalent to the nine phases through third-order invariants.
This observation suggests a certain duality between angles and phases.
Since the angles and phases are fourth- and fifth-order invariants, 
they are directly mapped onto each other through ninth-order invariants.

This structure becomes more transparent
by solving  Eqs.~(\ref{PXP}) and (\ref{DPX}) for $\Psi$. 
Eliminating X from the phases $\D$ and angles $\Phi$, 
one obtains the following relation
\begin{align}
\D -\Phi  = \Psi T + T \Psi - \Psi \, .   
\end{align}
The right-hand side becomes a ninth-order invariant, consisting of six $V_{\alpha i}$ and three $V_{\alpha i}^{*}$.

Alternatively, by adding $2 \Psi$ to both sides, one obtains the ninth-order invariant that involves all  $V_{\alpha i}$ without complex conjugates.
\begin{align}
\D -\Phi + 2 \Psi = (1 + T + T^{2} ) \Psi 
= 
\begin{pmatrix}
1 &1 & 1 \\
1 &1 & 1 \\
1 &1 & 1 \\
\end{pmatrix} (\psi_{1} + \psi_{2} + \psi_{3}) \equiv \Pi' . 
\end{align}
The right-hand side reduces to Eqs.~(\ref{detV3}) and (\ref{24}).
This is a nontrivial prediction, which is confirmed explicitly
\begin{align}
& 
\begin{pmatrix}
 92.44^{\circ} & 115.91^{\circ} & 65.15^{\circ} \\
 157.51^{\circ} & 1.089^{\circ} & 114.89^{\circ} \\
 23.54^{\circ} & 156.49^{\circ} & 93.45^{\circ} \\
\end{pmatrix} 
-
\begin{pmatrix}
 1.054^{\circ} & 22.49^{\circ} & 156.46^{\circ} \\
 64.09^{\circ} & 92.40^{\circ} & 23.50^{\circ} \\
 114.85^{\circ} & 65.11^{\circ} & 0.0370^{\circ} \\
 \end{pmatrix} 
  + 2
\begin{pmatrix}
 1.052^{\circ} & 0.035^{\circ} & 92.40^{\circ} \\
 0.035^{\circ} & 92.40^{\circ} & 1.052^{\circ} \\
 92.40^{\circ} & 1.052^{\circ} & 0.035^{\circ} \\
\end{pmatrix} \nn
\\ & = 
\begin{pmatrix}
 93.49^{\circ} & 93.49^{\circ} & 93.49^{\circ} \\
 93.49^{\circ} & 93.49^{\circ} & 93.49^{\circ} \\
 93.49^{\circ} & 93.49^{\circ} & 93.49^{\circ} \\
\end{pmatrix} 
= 
\begin{pmatrix}
1 &1 & 1 \\
1 &1 & 1 \\
1 &1 & 1 \\
\end{pmatrix} 
\D - 
\begin{pmatrix}
1 &1 & 1 \\
1 &1 & 1 \\
1 &1 & 1 \\
\end{pmatrix} \pi  . 
\end{align}
The reason why the expression is written by the simple difference is that the angle matrix $\Phi$ and the phase matrix $\D$ are defined so as not to share any common matrix elements of $V$. 

The other half of the sum rules $\D - (T + T^{T}) \Phi = (T + T^{T} - 1 ) {\rm X}$ 
is also obtained from Eqs.~(\ref{PXP}) and (\ref{DPX}). 
This is simplified by using Eq.~(\ref{TPi}), 
\begin{align}
\D + \Phi  =
 (T + T^{T} - 1 ) {\rm X} +   (1 + T + T^{2}) \pi \, .
\end{align}
By adding 2X to both sides,
\begin{align}
\D + \Phi + 2 {\rm X} =
 (1 + T + T^{2} ) {\rm X} +  (1 + T + T^{2}) \pi =  \Pi '   \, .
\end{align}
Numerically, it is verified as
\begin{align}
& 
\begin{pmatrix}
 92.44^{\circ} & 115.91^{\circ} & 65.15^{\circ} \\
 157.51^{\circ} & 1.089^{\circ} & 114.89^{\circ} \\
 23.54^{\circ} & 156.49^{\circ} & 93.45^{\circ} \\
\end{pmatrix} 
+ 
\begin{pmatrix}
 1.054^{\circ} & 22.49^{\circ} & 156.46^{\circ} \\
 64.09^{\circ} & 92.40^{\circ} & 23.50^{\circ} \\
 114.85^{\circ} & 65.11^{\circ} & 0.0370^{\circ} \\
 \end{pmatrix} 
+ 2 
\begin{pmatrix}
- 0.0019^{\circ} &  - 22.45^{\circ} & -64.06^{\circ} \\
-64.06^{\circ} & - 0.0019^{\circ} & - 22.45^{\circ}  \\
 - 22.45^{\circ} & -64.06^{\circ}  & - 0.0019^{\circ} \\
\end{pmatrix}
\nn \\ &  = 
\begin{pmatrix}
 93.49^{\circ} & 93.49^{\circ} & 93.49^{\circ} \\
 93.49^{\circ} & 93.49^{\circ} & 93.49^{\circ} \\
 93.49^{\circ} & 93.49^{\circ} & 93.49^{\circ} \\
\end{pmatrix}  . 
\end{align}
These equations precisely reveal the underlying order between scattered and chaotic values of the CP phases, angles, and third-order invariants.

Ultimately, these two equations of the duality lead to 
\begin{align}
\Phi = \Psi - {\rm X} \, , ~~~ 
\D =  \Pi' - \Psi - {\rm X} \, . 
\end{align}
In other words, the CP phases are given by the arguments of the product of all matrix elements divided by two third-order invariants with odd and even permutations.
These expressions without the transfer matrix $T$
 are the most concise and transparent form. 
Since such dualities are expected to exist for other orders, 
 higher-order invariants can be concisely represented in terms of lower-order ones.

\section{Summary}

In this paper, we demonstrate that all nine angles of the unitarity triangles and CP phases in the nine different Euler-angle-like representations are expressed by products of third-order rephasing invariants $V_{\a i} V_{\b j} V_{\g k} / \det V$.   
The fourth-order angles and the fifth-order phases are composed of two and three invariants.
This fact shows that the fifth-order invariants are not genuinely irreducible. 
The sum rules relating these nine angles and phases are also decomposed into terms of the third-order  invariants. 

Furthermore, through ninth-order invariants, these angles and phases become equivalent, 
which can be regarded as a certain duality. 
For the phase matrix $\D$ and the angle matrix $\Phi$,
$\D \pm \Phi$ are expressed in terms of even-permutations X and odd-permutations $\Psi$ of third-order invariants.
As a result, these are represented by the two concise matrix equations 
$\Phi = \Psi - {\rm X}$ and $\D =  \Pi' - \Psi - {\rm X}$. 
This demonstrates that the CP phases and angles, appearing to be random and scattered, are described in a unified way using more fundamental invariants.

These results indicate that the third-order invariants play an essential role in the understanding of flavored CP violation.  
Similarly, sixth-, seventh-, and higher-order invariants can also be constructed, and simple sum rules
and dualities are expected among their arguments and lower-order invariants.  
The framework developed in this study complements the traditional Jarlskog invariants, offering a deeper structural understanding of CP violation.  
Moreover, this approach is not limited to the CKM matrix but can also be applied to the MNS matrix, 
because varying the values of the Majorana phases corresponds to rephasing transformations of the neutrinos and the value of $\d$ remains unaffected. 
Therefore, this methodology provides an effective analytical tool for precision measurements and new physics searches.

\section*{Acknowledgment}

The study is partly supported by the MEXT Leading Initiative for Excellent Young Researchers Grant Number JP2023L0013.


\end{document}